\documentclass{article}
\usepackage{spconf,amsmath,graphicx}

\usepackage{dcolumn}
\usepackage{amsmath,amsfonts}
\usepackage{bm}
\usepackage{multirow}
\usepackage{tabularx}
\usepackage{mathtools}
\usepackage{tikz}
\usepackage{amsmath}
\usepackage{amssymb}
\usepackage{xcolor,colortbl}
\definecolor{LightCyan}{rgb}{0.88,1,1}
\definecolor{Gray}{gray}{0.85}

\title{Hierarchical cross-modality knowledge transfer with sinkhorn attention for CTC-based ASR}
\name{Xugang Lu$^{1*}$, Peng Shen$^{1}$, Yu Tsao$^{2}$, Hisashi Kawai$^1$}
\address{1. National Institute of Information and Communications Technology, Japan\\
	2. Research Center for Information Technology Innovation, Academic Sinica, Taiwan}
\begin{document}
%
\maketitle
\begin{abstract}
Due to the modality discrepancy between textual and acoustic modeling, efficiently transferring linguistic knowledge from a pretrained language model (PLM) to acoustic encoding for automatic speech recognition (ASR) still remains a challenging task. In this study, we propose a cross-modality knowledge transfer (CMKT) learning framework in a temporal connectionist temporal classification (CTC) based ASR system where hierarchical acoustic alignments with the linguistic representation are applied. Additionally, we propose the use of Sinkhorn attention in cross-modality alignment process, where the transformer attention is a special case of this Sinkhorn attention process. The CMKT learning is supposed to compel the acoustic encoder to encode rich linguistic knowledge for ASR. On the AISHELL-1 dataset, with CTC greedy decoding for inference (without using any language model), we achieved state-of-the-art performance with 3.64\% and 3.94\% character error rates (CERs) for the development and test sets, which corresponding to relative improvements of 34.18\% and 34.88\% compared to the baseline CTC-ASR system, respectively.  
\end{abstract}
\begin{keywords}
Pretrained language model (PLM), Cross-modality alignment, sinkhorn attention, automatic speech recognition (ASR)
\end{keywords}
\section{Introduction}
\label{sec:intro}
Due to the non-autoregressive (NAR) decoding capability for fast and parallel inference, the temporal connectionist temporal classification (CTC)-based learning \cite{CTCASR} for automatic speech recognition (ASR) is one of the most attractive frameworks for end to end (E2E) ASR \cite{Li2022}. However, token independence assumption in CTC based learning makes it difficult for acoustic encoder to learn rich context dependent linguistic information. Leveraging a language model (LM), particularly a pretrained language models (PLM) to improve the ASR performance is a promising direction. In early studies \cite{Watanabe2017}, based on attention with encoder-decoder (AED) modeling, a hybrid CTC/AED-based ASR model framework was proposed to enhance linguistic information in acoustic encoder. With multi-task learning framework, several methods have been proposed to learn linguistic information by inserting linguistic knowledge in intermediate layers of acoustic encoders for ASR \cite{HierarchicalCTC,intermediateCTC}. In recent years, due to the success of self-supervised learning in feature exploration, knowledge transfer learning from both pretrained acoustic model (e.g., wav2vec2.0 \cite{wav2vec2.0}) and PLM (e.g., bidirectional encoder representation from transformers (BERT) \cite{BERT}) for ASR also have been proposed \cite{CIFBERT1,CTCBERT1,Cho2020,FNAR-BERT,wav2vecBERTSLT2022}. 

Although stacking text encoder of a PLM on top of the acoustic encoder could improve the ASR performance\cite{NARBERT}, it is preferred to transfer linguistic knowledge encoded in the PLM to acoustic encoding via cross-modal knowledge distillation (KD) \cite{FNAR-BERT,KuboICASSP2022,Futami2022,Choi2022,Higuchi2022}. However, in most studies, the KD learning is carried out on the probability logits of the acoustic model or on the last hidden layer of acoustic encoders \cite{FNAR-BERT,Futami2022,KuboICASSP2022} . This learning paradigm is based on a simple assumption that high abstract-level of acoustic feature corresponds to tokens of linguistic knowledge in a bottom-up based processing of speech. However, as studies revealed that the acoustic feature learning even in low-level should be guided with linguistic knowledge as a top-down attention process \cite{SPAtt}. In this study, we propose a novel cross-modality knowledge transfer (CMKT) learning framework for linguistic knowledge transfer from a PLM to acoustic encoding in CTC based ASR. Our main contributions are summarized as: 1. A hierarchical acoustic alignments with the linguistic latent representations from a PLM are applied for CMKT, even in low-level of acoustic features; 2. In cross-modality alignment, we propose to use the Sinkhorn attention \cite{SinkhornAtt,Sinkformer} for feature alignment where the transformer attention \cite{Transformer} is a special case of the iteration of Sinkhorn normalization process; 3. We implemented the CMKT learning algorithm in a CTC based ASR for acoustic encoding and confirmed its effectiveness via detailed experiments.       
\section{Proposed method}
\label{sec:proposed}
\begin{figure}[tb]
	\centering
	\includegraphics[width=8cm, height=5.5cm]{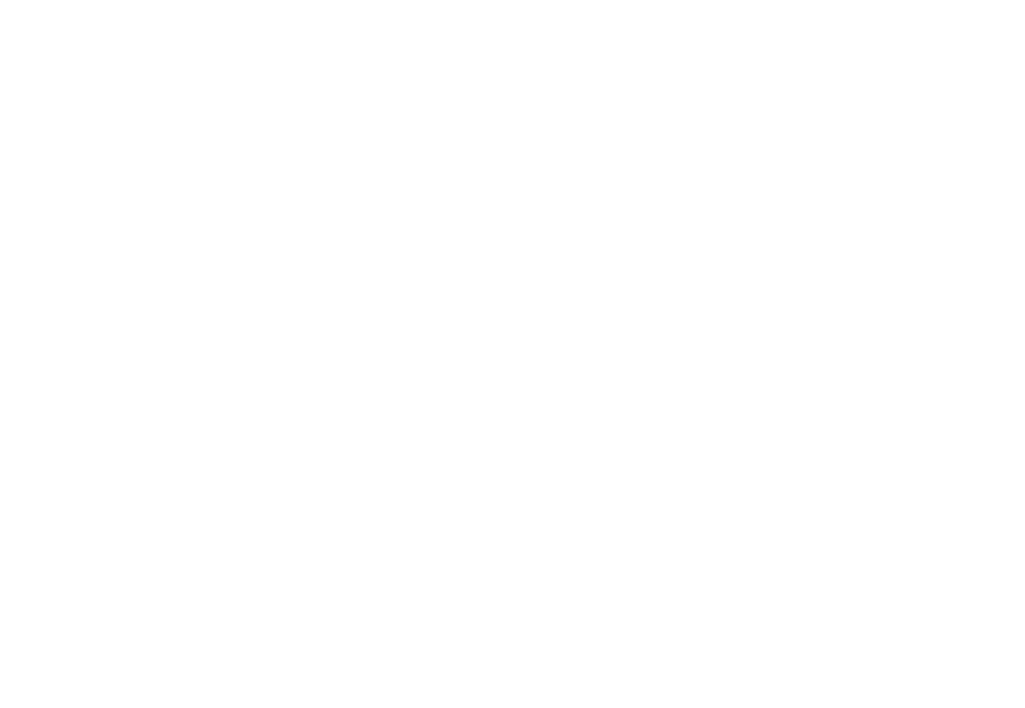}
	\caption{The proposed cross-modality knowledge transfer framework for CTC-based ASR.}
	\vspace{-1mm}
	\label{fig:frame1}
\end{figure}

\begin{figure}[tb]
	\centering
	\includegraphics[width=8cm, height=5.5cm]{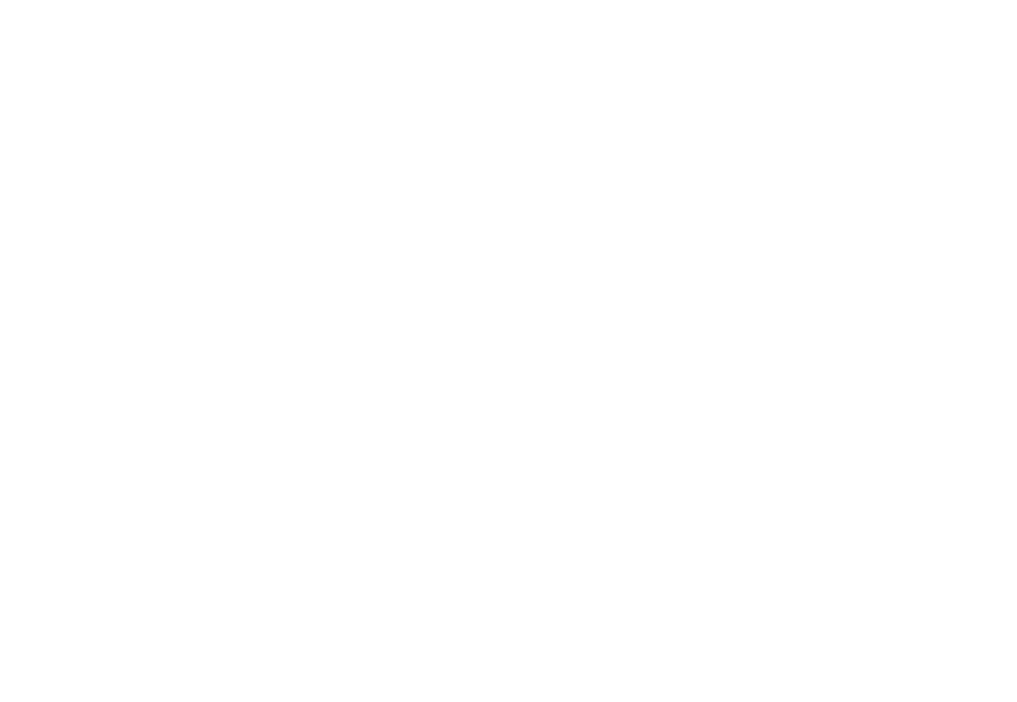}
	\caption{Adapter and Cross-modality matching modules which are shared by all encoder blocks.}
	\vspace{-2mm}
	\label{fig:frame2}
\end{figure}

The proposed model framework is illustrated in Fig. \ref{fig:frame1}, and the adapter and cross-modality matching modules in Fig. \ref{fig:frame1} are further explained in Fig. \ref{fig:frame2}. In these figures, `FC1', `FC2', `FC3' are linear transforms of fully-connected layers, `LN' denotes layer-normalization, and `CM-encoder' represents cross-modality encoder. In the follows, we will explain the process of each module. 
\subsection{Features from acoustic and textual modalities}
In acoustic modality of Fig. \ref{fig:frame1}, the process in `Subsampling' module and position encoding of acoustic sequence (${\rm PE}_{\rm A}$) are used to extract the initial input feature as ${\bf G}_0$. Then the output from the $i$-th acoustic encoder block is represented as:
\begin{equation}
	{\bf G}_i  = {{\rm Encoder}_i}\left( {{\bf G}_{i - 1} } \right) \in \mathbb{R}^{l_a  \times d_a }
	\label{eq:conformer}
\end{equation}
where $i$ takes values from $1$ to $M_a$, with $M_a$ representing the total number of encoder blocks, $l_a$ and $d_a$ are length (temporal dimension) and feature dimension, respectively. 
In adapter process, a linear transform FC2 is applied for feature dimension matching between acoustic and textual modalities (in Fig. \ref{fig:frame2}):
\begin{equation}
	{\bf H}_i  = {\rm FC}_{\rm 2} \left( {{\bf G}_i } \right) \in R^{l_a  \times d_t } 
	\label{eq:FC2}
\end{equation}
In this ${\bf H}_i$, the feature dimension is $d_t$ corresponding to textual feature dimension, and $i$ is the encoder block index.

The initial textual feature is obtained from token embedding `EMB' and position encoding of text `${\rm PE}_{\rm T}$' as ${\bf Z}_0 \in R^{l_t  \times d_t}$ with sequence length $l_t$ and feature dimension $d_t$. This ${\bf Z}_0$, together with representations from acoustic modality is transformed by a sequence of cross-modality encoders (as CM-encoder in Fig. \ref{fig:frame2}). For an acoustic representation ${\bf H}$ from Eq. (\ref{eq:FC2}) (encoder block index is omitted for easy explanation), the transform in each CM-encoder is formulated as: 
\begin{equation}
	{\bf Z}_i  = f_i ({\bf Z}_{i - 1} ,{\bf H}),
\end{equation} 
where $i$ is the index of textual encoder block with values from $1$ to $M_t$. In each CM-encoder, there is a sequential process with modality feature transform, layer-normalization, and feed forward transform as:
\begin{equation}
	{\bf Z}_{{\bf H} \to {\bf Z}_{i - 1} }  = {\rm OT}\left( {{\bf H} \to {\bf Z}_{i - 1} } \right)  
	\label{eq:cfea}
\end{equation} 
\begin{equation}
	\begin{array}{l}
		\begin{aligned}					
		{\bf \hat Z}_{i - 1} {\rm  = LN}\left( {{\bf Z}_{i - 1}  + {\bf Z}_{{\bf H} \to {\bf Z}_{i - 1} } } \right) \\ 
		{\bf Z}_i  = {\rm LN}\left( {{\bf \hat Z}_{i - 1}  + {\rm FC}\left( {{\bf \hat Z}_{i - 1} } \right)} \right) \\  
		\end{aligned}
	\end{array}
	\label{eq:cfea1}
\end{equation} 
In Eq. (\ref{eq:cfea}), ${\bf Z}_{{\bf H} \to {\bf Z}_{i - 1} }$ is a transported representation (from acoustic modality to textual modality), and $\rm OT(.)$ denotes optimal transport (OT). As it is showed that the acoustic feature ${\bf H}  \in \mathbb{R}^{l_a  \times d_a }$, and the textual feature ${\bf Z}_{i - 1}  \in \mathbb{R}^{l_t  \times d_t } $ are two modalities with different lengths. After OT, the representation ${\bf Z}_{{\bf H} \to {\bf Z}_{i - 1} }$ keeps the same dimensions of that of ${\bf Z}_{i - 1}$. The final output of the cross-modality matching is represented as ${\bf Z}_{M_t}$. For transferring linguistic knowledge from the BERT, we suppose that this ${\bf Z}_{M_t}$ should approximate a target textual representation which is provided by the pretrained BERT model as:
\begin{equation}
	\begin{array}{l}
		{\bf y}_{{\rm token}}  = {\rm Tokenizer}\left( {\bf y} \right); {\bf \tilde Z}_0  = [{\rm CLS, }{\bf y}_{{\rm token}} ,{\rm SEP]} \\ 
		{\bf \tilde Z}_i  = {\rm BERT}_i \left( {{\bf \tilde Z}_{i - 1} } \right) \in \mathbb{R}^{l_t  \times d_t } \\ 
	\end{array}	
	\label{eq:bert}
\end{equation}
where `${\rm BERT}_{i}$' is the $i$-th transformer encoder layer of BERT model, $i$ takes values from $1$ to $M_b$, with $M_b$ representing the total number of BERT encoder layers. `$\rm Tokenizer$' is a process to convert standard text to word piece based tokens \cite{BERT}. Token symbols `CLS` and `SEP` represent the start and end of an input sequence. In model learning stage, for linguistic knowledge transfer, the loss function is defined as cross-modal alignment loss by:
\begin{equation}
	L_{{\rm align}}  = \sum\limits_{j = 2}^{l_t  - 1} {1 - \cos \left( {{\bf z}_{j,:} ,{\bf \tilde z}_{j,:} } \right)},
	\label{eq:align}
\end{equation}
where ${\bf z}_{j,:}$ and ${\bf \tilde z}_{j,:}$ are row vectors of feature matrices ${\bf Z}_{M_t}$ and ${\bf \tilde Z}_{i}$ from `${\rm BERT}_{i}$', respectively. In this formulation, the sum ranges from 2 to $l_t-1$ in order to exclude the `[CLS]' and `[SEP]' tokens from the loss estimation (refer to Eq. (\ref{eq:bert}) in text encoding). 
\subsection{Sinkhorn attention for CMKT learning}
In Eq. (\ref{eq:cfea}), the solution is represented as (index subscript is omitted for easy explanation):
\begin{equation}	
	{\bf Z}_{{\bf H} \to {\bf Z}}  = \hat \gamma  \times {\bf H} \in R^{l_t  \times d_t }   		
\end{equation}
where $\hat \gamma$ is a transport coupling matrix based on minimizing an entropy regularized OT (EOT) $L_{EOT} ({\bf H},{\bf Z} )$ as:
\begin{equation}
	\hat \gamma  = \mathop {\arg \min }\limits_{\gamma  \in \prod {({\bf H},{\bf Z})} } L_{EOT} ({\bf H},{\bf Z})
	\label{eq:gammamin}
\end{equation} 
And the objective function $L_{EOT} ({\bf H},{\bf Z} )$ is defined as: 
\begin{equation}
	L_{EOT} ({\bf H},{\bf Z})\mathop  = \limits^\Delta  \sum\limits_{i,j} {\gamma _{i,j} C_{i,j}  + \alpha \gamma _{i,j} \log \gamma _{i,j} }, 
	\label{eq:EOTloss}  
\end{equation}
where $\alpha$ is a regularization coefficient, $\gamma _{i,j}$ and $C_{i,j}$ are elements of transport coupling $\gamma$ and cost matrices $C$. The solution of Eq. (\ref{eq:gammamin}) can be implemented as an iteration of Sinkhorn projections as \cite{SinkhornAtt,VillanoBook}:
\begin{equation}
		\gamma ^0  = \exp \left( { - \frac{1}{\alpha }C} \right), \gamma ^{k + 1}  = F_c \left( {F_r \left( {\gamma ^k } \right)} \right), 		
	\label{eq:scaling}
\end{equation}
where $F_c(.)$ and $F_r(.)$ are column- and row-wise normalization operators, respectively. 
In real applications, in Eq. (\ref{eq:scaling}), only a few iterations are enough to obtain fairly well results (in our experiments, 3 times of iterations were set).

In iteration process of Sinkhorn attention, the row-wise normalization $F_r(.)$ can be formulated as:
\begin{equation}
	F_r \left( \gamma  \right)  = \frac{{\gamma }}{{\sum\limits_j {\gamma _{i,j} } }} = {\rm softmax}\left( {{\rm FC}\left( C \right)} \right),
	\label{eq:normsoftmax}
\end{equation}
where `FC' is a linear transform. When choosing negative inner product as the cost function for $C$, Eq. (\ref{eq:normsoftmax}) is further cast to:
\begin{equation}
	F_r \left( \gamma  \right)  = {\rm softmax}\left( {{\bf ZW}_{\bf Z} \left( {{\bf HW}_{\bf H} } \right)^T } \right),
	\label{eq:gamma}
\end{equation}
where ${\bf W}_{\bf Z}$ and ${\bf W}_{\bf H}$ are feature transform matrices for acoustic and textual representations, respectively. From this equation, we can see that the transformer attention \cite{Transformer} can be regarded as a special case of the Sinkhorn attention with proper chosen of linear transforms and cost functions which also have been studied in \cite{Sinkformer}. 

\subsection{Loss function in CMKT learning}
For transferring back linguistic information in acoustic encoding, the following transforms are designed as indicated in Fig. \ref{fig:frame2}:
\begin{equation}
	\begin{array}{l}
		{{\bf \hat H}_i} = {\rm FC3}\left( {{\rm LN}\left( {\bf H}_i \right)} \right) \in R^{l_a  \times d_a }  \\ 
		{\bf H}_i^{a,t}  = {\bf G}_i  + {\rm LN}\left( {{\bf \hat H}_i} \right) \\ 
	\end{array}
	\label{eq:adapter}
\end{equation}
Based on this new representation ${\bf H}^{a,t}$ which is supposed to encode both acoustic and linguistic information, the final probability prediction for ASR is formulated as:
\begin{equation}
	{\bf \tilde P} = {\rm Softmax}\left( {{\rm FC1}\left({\bf H}_{M_a }^{a,t} \right)} \right)
	\label{eq:softmaxadd}
\end{equation}
In training with CMKT, the total loss is defined as:
\begin{equation}
	L\mathop  = \limits^\Delta  \lambda .L_{{\rm CTC}} ({\bf \tilde P},{\bf y}_{{\rm token}} ) + (1 - \lambda ).w.\sum\limits_{i = 1}^{M_a } {(L_{{\rm align}}^i  + L_{{\rm EOT}}^i )},   
	\label{eq:totalloss} 
\end{equation}
where $L_{{\rm CTC}} ({\bf \tilde P},{\bf y}_{{\rm token}} )$ is CTC loss, ${L_{{\rm align}}^i }$ and ${L_{{\rm EOT}}^i }$ are cross-modality alignment loss and OT loss collected from the hierarchical encoder block indexed by $i$ as defined in Eqs. (\ref{eq:align}) and (\ref{eq:EOTloss}), respectively. $\lambda$ is a trade off parameter, $w$ is a parameter to scale the alignment loss. After the model is trained, only the left branch of Fig. \ref{fig:frame1} is kept for ASR inference.  


\section{Experiments}
\label{sec:exp}
We carried out experiments on an open source Mandarin speech corpus AISHELL-1 which includes speech recorded from 400 speakers \cite{AISHELL1}. Three data sets are included: a training set with 340 speakers (150 hours), a development (or validation) set with 40 speakers (10 hours), and a test set with 20 speakers (5 hours). In training, data augmentation with speed perturbation (with factors of 0.9 and 1.1) was applied \cite{AISHELL1}. 80-dimensional log Mel-filter bank features together with 3-dimensional fundamental frequency related features (F0, delta F0 and delta delta F0) are used as raw input feature, and they were extracted with a 25ms window size and a 10ms shift. 
\subsection{Parameter settings}
In acoustic modality, the convolutional block in CNN subsampling module is with 256 channels, kernel size 3, stride 2, and ReLU activation function. Conformer based acoustic encoder \cite{conformer2020} is used. In each conformer block, the convolution is with kernel size of 15, attention dimension is $d_a=256$, attention head is 4, and the dimension of FFN layer is 2048. The BERT of `bert-base-chinese' from huggingface is used as the PLM \cite{Huggingface}. In this BERT model, there are $M_b=12$ transformer encoders, token size is 21128, and text feature dimension is $d_t=768$. For reducing calculation redundancy, the CMTK was carried out on every three layers of the acoustic encoders. Several other hyper-parameters are fixed and set as: EOT regularization parameter $\alpha=1.0$ in Eqs. (\ref{eq:EOTloss}) and (\ref{eq:scaling}), scale parameter $w=1.0$ and alignment trade off parameter $\lambda=0.3$ in Eq. (\ref{eq:totalloss}). In optimization, Adam optimizer \cite{Adam} is used with a learning rate (initial with 0.001) schedule with 20,000 warm-up steps. The model was trained for 130 epochs, and the final model used for evaluation was obtained by averaging models from the last 10 epochs. The performance was evaluated based on character error rate (CER).   
\subsection{Results}
\label{sec:results}
The model is trained by fixing hyper-parameter settings of acoustic encoder layers $M_a=16$, textual encoder layers $M_t=5$, and three times of iteration in Sinkhorn attention. After the model is learned, CTC greedy search based decoding is used for recognition where only the components in acoustic modality is used. The results are showed in table \ref{tab:tab1}. For comparison, the results of baseline system and several state-of-the-art systems which integrate BERT for linguistic knowledge transfer are also showed in Table \ref{tab:tab1}. 
\begin{table}[tb]
	\centering
	\caption{ASR performance on AISHELL-1 coprus, CER (\%).}
	\begin{tabular}{|c||c||c|}
		\hline
		Methods &dev set &test set\\
		\hline
		Conformer-CTC (Baseline)  &5.53 &6.05 \\
		\hline	
		\hline	
		Conformer-CTC/AED (\cite{Watanabe2017})  &4.61 &5.06 \\						
		\hline
		NAR-BERT-ASR (\cite{NARBERT}) &4.90 &5.50 \\
		\hline
		LASO with BERT (\cite{FNAR-BERT}) &5.20 &5.80 \\
		\hline
		KT-RL-ATT (\cite{CTCBERT1}) &4.38 &4.73 \\
		\hline
		Wav2vec-BERT (\cite{wav2vecBERTSLT2022}) &4.10 &4.39 \\
		\hline
		Last-CMKT (proposed)  &\textbf{4.05} &\textbf{4.40} \\	
		\hline
		Hierarchical-CMKT (proposed)  &\textbf{3.64} &\textbf{3.94} \\			
		\hline		
	\end{tabular}
	\label{tab:tab1}
\end{table}
In this table, the `Conformer-CTC' is the baseline system. `Conformer-CTC/AED' denotes a hybrid CTC/AED ASR system as proposed in\cite{Watanabe2017}. `KT-RL-ATT' \cite{CTCBERT1}, and `Wav2vec-BERT' \cite{wav2vecBERTSLT2022} took pretrained acoustic model (from wav2vec2.0 \cite{wav2vec2.0}) and BERT for knowledge transfer. The two methods with `Last-CMKT' or `Hierarchical-CMKT' represents that our proposed CMKT was applied on the last hidden layer or on hierarchical of the acoustic encoders. From this table, we can see that our proposed CMKT yields competitive results. In particular, hierarchical CMKT achieved state of the art performance which suggested that linguistic knowledge transfer should be on both high and low-level of acoustic abstractions in order to improve ASR performance. 
\vspace{-3mm}
\subsection{Ablation study}
In this section, we figure out several important factors which affect the ASR performance in CMKT learning. 
\subsubsection{How many textual encoder layers are sufficient?}
In textual encoder, several `CM-encoder' layers are used to explore linguistic information with reference to features from acoustic encoder (refer to Fig. \ref{fig:frame2}). With target linguistic representation from BERT as a supervision signal, the textual encoder could explore textual information from both textual and acoustic modalities. We did experiments with different number of CM-encoder layers, and showed results in table \ref{tab:tab3}. 
\begin{table}[tb]
	\centering
	\caption{ASR performance with different number of CM-encoder layers, CER (\%).}
	\begin{tabular}{|c||c||c|}
		\hline
		\# CM-encoder layers&dev set &test set\\
		\hline
		$M_t=1$  &5.11 &5.60 \\
		\hline
		$M_t=3$  &3.69 &4.05 \\
		\hline			
		$M_t=5$  &\textbf{3.64} &\textbf{3.94} \\						
		\hline
		$M_t=7$ &3.66 &3.99 \\
		\hline		
	\end{tabular}
	\vspace{-3mm}
	\label{tab:tab3}
\end{table}
From this table, we can see that it is necessary to increase the number of CM-encoder layers for the purpose of increasing textual encoder's capability to fully explore the information from textual and acoustic modalities. 
\subsubsection{Is the adapter necessary?}
The adapter is a connection to pass acoustic information to text modality in CMKT learning, and transfer back the textual information conditioned on acoustic representations. Two experimental conditions are examined, i.e., Condition1: the adapter module is integrated in acoustic modality but no CMKT learning is performed (i.e., cut-off the connection to textual modality); Condition 2: the adapter is connected to textual modality with CMKT learning but it is not connected back to acoustic encoder (i.e., cut-off the FC3 link to acoustic modality in Fig. \ref{fig:frame2}). The results are shown in table \ref{tab:tab4}.
\begin{table}[tb]
	\centering
	\caption{ASR performance with and without adapter connections in CMKT learning, CER (\%).}
	\begin{tabular}{|c||c||c|}
		\hline
		Adapter connections&dev set &test set\\
		\hline
		Condition 1  &5.25 &5.77 \\
		\hline
		Condition 2  &4.20 &4.54 \\
		\hline					
	\end{tabular}
	\vspace{-3mm}
	\label{tab:tab4}
\end{table}
From this table, we can see that the CMKT is the most important part in the proposed framework, and adapter with connections to both acoustic and textual modalities are also necessary.

\section{Conclusion}
\label{sec:conclusion}
In this study, we propose a novel hierarchical CMKT learning approach to enhance CTC-based ASR by harnessing linguistic representations encoded in a PLM model. CMKT learning involves transferring linguistic knowledge at both high and low levels of acoustic representations. In CMKT, we design Sinkhorn attention with just a few iterations to align cross-modal features. Using this alignment, the textual encoder can extract information from both textual and acoustic modalities to approximate the target linguistic representations encoded in BERT. By using an adapter that connects both acoustic and textual modalities, we efficiently transfer linguistic knowledge to the acoustic encoder. Our experiments confirm the effectiveness of the proposed CMKT learning framework. 

The capacity of the proposed CMKT learning framework has not been fully explored. For example, questions remain regarding the integration of latent representations from the BERT model in CMKT learning and the adjustment of various hyperparameters in objective functions, especially concerning the Sinkhorn attention. In our future work, we will delve deeper into the potential of this learning framework through rigorous experimentation.   

\vfill\pagebreak

\bibliographystyle{IEEEbib}

\begin{thebibliography}{1}
	\bibitem{CTCASR}
	A. Graves, and N. Jaitly, ``Towards end to-end speech recognition with recurrent neural networks," in \emph{Proc. ICML}, pp. 1764–1772, 2014.
	
	\bibitem{Li2022}
	J. Li, ``Recent advances in end-to-end automatic speech recognition," \emph{APSIPA Transactions on Signal and Information Processing}, DOI 10.1561/116.00000050, 2022.
		
	\bibitem{Watanabe2017}
	S. Watanabe, T. Hori, S. Kim, J. R. Hershey and T. Hayashi, ``Hybrid CTC/Attention Architecture for End-to-End Speech Recognition," \emph{IEEE Journal of Selected Topics in Signal Processing}, vol. 11, no. 8, pp. 1240-1253, 2017.
	
	
	\bibitem{HierarchicalCTC}
	Y. Higuchi, K. Karube, T. Ogawa, T. Kobayashi, ``Hierarchical conditional end-to-end asr with ctc and multi-granular subword units," in \emph{Proc. of ICASSP}, pp. 7797-7801, 2022.
	
	\bibitem{intermediateCTC}
	Y. Fujita, T. Komatsu, and Y. Kida, ``Alternate Intermediate Conditioning with Syllable-Level and Character-Level Targets for Japanese ASR," in \emph{Proc. of SLT}, pp. 76-83, 2022.
	
	
	\bibitem{wav2vec2.0}
	A. Baevski, Y. Zhou, A. Mohamed, and M. Auli, ``Wav2vec 2.0: A framework for self-supervised learning of speech representations," in \emph{Proc. of NeurIPS}, 2020. 
	
		
	\bibitem{BERT}
	J. Devlin, M. Chang, K. Lee, and K. Toutanova, ``Bert: Pretraining of deep bidirectional transformers for language understanding," \emph{arXiv preprint} arXiv:1810.04805, 2018.	
	
	\bibitem{CIFBERT1}
	M. Han, F. Chen, J. Shi, S. Xu, B. Xu, ``Knowledge Transfer from Pre-trained Language Models to Cif-based Speech Recognizers via Hierarchical Distillation," \emph{arXiv preprint} arXiv:2301.13003, 2023.
	
	\bibitem{CTCBERT1}
	K. Deng, S. Cao, Y. Zhang, L. Ma, G. Cheng, J. Xu, P. Zhang, ``Improving CTC-Based Speech Recognition Via Knowledge Transferring from Pre-Trained Language Models," in \emph{Proc. of ICASSP}, pp. 8517-8521, 2022.
	
	\bibitem{Cho2020}
	W. Cho, D. Kwak, J. Yoon, N. Kim, ``Speech to Text Adaptation: Towards an Efficient Cross-Modal Distillation," in \emph{Proc. of INTERSPEECH}, pp. 896-900, 2020.
	
	\bibitem{FNAR-BERT}
	Y. Bai, J. Yi, J. Tao, Z. Tian, Z. Wen and S. Zhang, "Fast End-to-End Speech Recognition Via Non-Autoregressive Models and Cross-Modal Knowledge Transferring From BERT," \emph{IEEE/ACM Transactions on Audio, Speech, and Language Processing}, vol. 29, pp. 1897-1911, 2021.
	
	\bibitem{wav2vecBERTSLT2022}
	K. Lu and K. Chen, ``A Context-aware Knowledge Transferring Strategy for CTC-based ASR," in \emph{Proc. of SLT}, pp. 60-67, 2022.
		
	\bibitem{NARBERT}
	F. Yu, K. Chen, and K. Lu, ``Non-autoregressive ASR Modeling using Pre-trained Language Models for Chinese Speech Recognition," \emph{IEEE/ACM Transactions on Audio, Speech, and Language Processing}, vol. 30, pp. 1474-1482, 2022
	
	\bibitem{KuboICASSP2022}
	Y. Kubo, S. Karita, M. Bacchiani, ``Knowledge Transfer from Large-Scale Pretrained Language Models to End-To-End Speech Recognizers," in \emph{Proc. of ICASSP}, pp. 8512-8516, 2022.
	
	\bibitem{Futami2022}
	H. Futami, H. Inaguma, M. Mimura, S. Sakai, T. Kawahara, ``Distilling the Knowledge of BERT for CTC-based ASR," CoRR abs/2209.02030, 2022.

    \bibitem{Choi2022}
    K. Choi, H. Park, ``Distilling a Pretrained Language Model to a Multilingual ASR Model," in \emph{Proc. of INTERSPEECH}, pp. 2203-2207, 2022.
    
    \bibitem{Higuchi2022}
    Y. Higuchi, T. Ogawa, T. Kobayashi, S. Watanabe, ``BECTRA: Transducer-based End-to-End ASR with BERT-Enhanced Encoder," CoRR abs/2211.00792, 2022.
    	
    \bibitem{SPAtt}
    C. Brodbeck, S. Bhattasali, A. Heredia, P. Resnik, J. Simon, E. Lau, ``Parallel processing in speech perception with local and global representations of linguistic context," \emph{Elife}, doi: 10.7554 /eLife .72056, 2022.
    
	\bibitem{SinkhornAtt}
	Y. Tay, D. Bahri, L. Yang, D. Metzler, D. Juan, ``Sparse Sinkhorn Attention," in \emph{Proc. of ICML}, pp. 9438-9447, 2020.
	
	\bibitem{Sinkformer}
	M. Sander, P. Ablin, M. Blondel, G. Peyre, ``Sinkformers: Transformers with Doubly Stochastic Attention," in \emph{Proc. of AISTATS}, pp. 3515-3530, 2022.
	
	\bibitem{Transformer}
	A. Vaswani, N. Shazeer, N. Parmar, J. Uszkoreit, L. Jones, A. Gomez, L. Kaiser, and I. Polosukhin, ``Attention is all you need," in \emph{Proc. of NIPS}, pp. 5998-6008, 2017.		
	
	\bibitem{VillanoBook}
	C. Villani, Optimal transport: old and new, volume 338. Springer, 2009
	
	\bibitem{AISHELL1} 
	H. Bu, J. Du, X. Na, B. Wu, and H. Zheng, ``AIShell-1: An open-source mandarin speech corpus and a speech recognition baseline,” in \emph{Proc. of COCOSDA}, pp. 1-5, 2017.
	
	\bibitem{conformer2020}
	A. Gulati, J. Qin, C. Chiu, et al., ``Conformer: Convolution augmented transformer for speech recognition," \emph{arXiv preprint} arXiv:2005.08100, 2020
	
	\bibitem{Huggingface}
	https://huggingface.co/
	
	\bibitem{Adam}
	D. Kingma, J. Ba, ``Adam: A Method for Stochastic Optimization," in \emph{Proc. of ICLR}, 2015.
	
\end{thebibliography}

\end{document}